\documentstyle[prb,aps,twocolumn,psfig]{revtex}
\begin{document}
\draft

\title{Masses of composite fermions carrying two
and four flux quanta: Differences and similarities}
\author{Xiaomin Zu, Kwon Park, and Jainendra K. Jain}
\address{Department of Physics, 104 Davey Laboratory,
The Pennsylvania State University,
University Park, Pennsylvania 16802}
\date{\today}

\maketitle

\begin{abstract}

This study provides a theoretical rationalization for the 
intriguing experimental observation regarding the equality of the 
``normalized" masses of composite fermions carrying two and four flux quanta, and 
also demonstrates that the mass of the latter type of composite fermion has 
a substantial filling factor dependence in the filling factor range
$4/17 > \nu > 1/5$, in agreement with experiment, originating from the relatively 
strong inter-composite fermion interactions here.

\end{abstract}


This work has been motivated by certain perplexing experimental results 
obtained recently by Yeh {\em et al.} \cite {Yeh} and Pan {\em et al.} 
\cite {Pan} regarding the composite fermion mass. 
Composite fermions are formed when electrons confined to two dimensions are exposed
to a strong magnetic field, since they best screen the repulsive Coulomb interaction
by effectively absorbing an even number of magnetic flux quanta each to turn into
composite fermions \cite {Jain,book1,book2}.  Numerous  properties of composite
fermions have been investigated over the past decade;
in particular, their Fermi sea, Shubnikov-de Haas oscillations, cyclotron
orbits and quantized Landau levels have been observed, and their
charge, spin, statistics, mass, g factor and thermopower
have been measured \cite{Stormer}.  The subject of this article is the masses of 
composite fermions carrying two and four flux quanta, denoted by $^2$CFs and $^4$CFs,
respectively.  The authors of Refs.~\onlinecite{Yeh} and \onlinecite {Pan}
define a dimensionless ``normalized" mass $m^*_{nor}=m^*/(m_e \sqrt{B[T]})$
where $m^*$ is the composite fermion (CF) mass, $m_e$ is the electron mass in vacuum
and $B[T]$ is the magnetic field quoted in Tesla.
(The electron mass in vacuum is chosen only as a convenient reference;
the CF mass is not related to it in any way.)
One unexpected outcome of the experiments was that the mass of  
composite fermion carrying four flux quanta at $\nu={n}/{(4n+1)}$
behaves qualitatively differently than the mass of composite fermion 
carrying two flux quanta at $\nu=n/(2n+1)$.  Specifically, whereas for
the latter the mass is approximately constant sufficiently far from the composite
fermion Fermi sea at $\nu=1/2$,  for the former the mass 
increases substantially as the filling factor decreases from $\nu={4}/{17}$ to 
$\nu={1}/{5}$.  This behavior was attributed to the presence of the 
localized Wigner crystal state at still smaller filling factors.
However, one may ask:  Is this effect intrinsic or driven by disorder?  If
intrinsic, can our present theoretical understanding account for it?  
What is its physical origin? Another remarkable finding 
was that the values of $m^*_{nor}$'s for composite fermions carrying two and 
four flux quanta are quite close in regions where they do not have strong filling 
factor dependence.  This was entirely unforeseen and rather  
difficult to justify in general terms; in fact, for a strictly two 
dimensional system, the $m^*_{nor}$ at $\nu={1}/{3}$ and
$\nu={1}/{5}$ can be determined from the known gaps to differ by a factor
of 2.5.  These experimental results thus pose well defined questions for the 
theory to explain.  It is clear that the explanation will not involve any 
universal symmetry principles, but rather a detailed, 
quantitative understanding of the physics.

In the limit of very
large magnetic fields, it is a good approximation (made throughout this work) to
restrict the Hilbert space of states to the lowest Landau level of electrons.
The kinetic energy degree of freedom of the electron is then quenched, 
and the defining problem contains no mass parameter.  The physical picture 
that we wish to present here is as follows.  When electrons transform into 
composite fermions, the interaction energy of electrons is converted into 
two parts: the kinetic energy of composite fermions, which defines a 
``bare" mass for the composite fermion,  and a residual interaction
between composite fermions.  The experimentally measured mass is of course 
the fully renormalized CF mass. 
When the inter-CF interaction is weak, the measured mass 
is roughly equal to the bare mass, and therefore  
approximately constant as a function of the filling
factor.  However, when the inter-CF interaction is non-negligible,  
the measured mass can be substantially different from the bare mass 
and also have significant filling factor dependence.  
Both the bare mass and its renormalization are formally of the same
order, originating from the same underlying electron-electron interaction, 
and it is not obvious {\em a priori} how to separate them theoretically.
Nonetheless, we will see that, on the basis of general considerations, 
it is possible to deduce theoretically a bare mass 
for composite fermion.  We find that (i) the
regions where the composite fermion mass has significant filling factor dependence
coincide with the regions where the residual interactions are strong; and 
(ii) the bare masses of composite fermions carrying two and four flux 
quanta are approximately equal.

The mass of composite fermions carrying $2p$ flux qauanta ($^{2p}$CFs) at
$\nu=n/(2pn+1)$ will be 
defined by equating the activation energy $\Delta$ measured in transport
experiments to an effective cyclotron energy of the composite fermion \cite{HLR}:
$\Delta[{n}/(2pn+1)] =\hbar {eB^*}/{m^*c}=
{\hbar^2}/[(2pn+1)m^* l_0^2]$ 
where $l_0=\sqrt{\hbar c /eB}$ is the magnetic length, and $B^*$ is the effective  
magnetic field for composite fermion, given
by $B^*={B}/{(2pn+1)}$ at $\nu={n}/{(2pn+1)}$.
On the other hand, since the gaps are determined entirely
by the Coulomb interaction, the only energy in the lowest Landau level 
constrained problem, they must be proportional to ${e^2}/{\epsilon l_0}$, 
$\epsilon$ being the dielectric constant of the background material.
Defining $\Delta=\delta\; ({e^2}/{\epsilon l_0}),$ one obtains 
$m^*={\hbar^2\epsilon}/[(2pn+1)\delta e^2l_0].$ 
For GaAs, with $\epsilon=12.8$ and band mass $=0.07 m_e$,  
the dimensionless mass is related to $\delta$ as  
$m^*_{nor}={0.0264}/{[(2pn+1){\delta}]}.$

Experimentally, rather than using the gaps directly, 
the mass of composite fermion is often determined 
from an analysis of the temperature dependence of the 
Shubnikov-de Haas (SdH) oscillations of composite fermions, 
or from the slope of the activation gaps plotted as a function of the effective
magnetic field, with the expectation that the mass thus obtained is less 
affected by the disorder, and is consequently closer to the intrinsic mass than 
the mass obtained directly from the experimental activation gaps.  
Incidentally, a different  mass of composite fermion is the ``polarization 
mass," \cite{Park98} determined from the values of the Zeeman energy
at which transitions between ground states of different polarizations
take place;  its normalized value for $^2$CFs was estimated 
theoretically \cite{Park98} to be $\approx 0.6$,
in good agreement with subsequent experiments \cite {Melinte,Kukushkin}.

To investigate the findings of Refs.~\onlinecite{Yeh} and 
\onlinecite {Pan}, we have 
computed the gaps for fractional quantum Hall states at $\nu={n}/{(4n+1)}$
in a Monte Carlo scheme using the Jain wave functions\cite{Jain}, and employing
the lowest-Landau-level projection technique described in Ref.~\onlinecite{JK}.  
In each case, the thermodynamic limit for the gap is obtained by a consideration 
of systems with up to 100 particles.  Up to 10 million 
Monte Carlo steps are performed for each energy (20 million for relatively large
systems, containing 70-100 particles); the error in the 
thermodynamic gap is a combination of
the uncertainty in the gaps of finite systems and the uncertainty in 
their extrapolation to the thermodynamic limit.  The wave function in the transverse
direction, needed for the determination of the effective two-dimensional
inter-electron interaction, 
is evaluated as a function of the two-dimensional density  ($\rho$) in a
self-consistent local density approximation (LDA), for a heterojunction sample.
$V_{eff}(r)$ differs from the Coulomb 
interaction at short distances, causing a reduction in the value of the gap.
Disorder is known to further reduce the gap, but in
the absence of a reliable theoretical approach to deal with it, it  
has been set to zero in the following.  Landau level (LL) 
mixing is also neglected, but
it is known to make only a slight correction to the value of the gap after the 
non-zero thickness is incorporated \cite {Bonesteel}.  
There is no adjustable parameter in the calculations.
For further details, the reader 
is referred to Ref.~\onlinecite{Park99} and the references therein.

\begin{figure}
\centerline{\psfig{figure=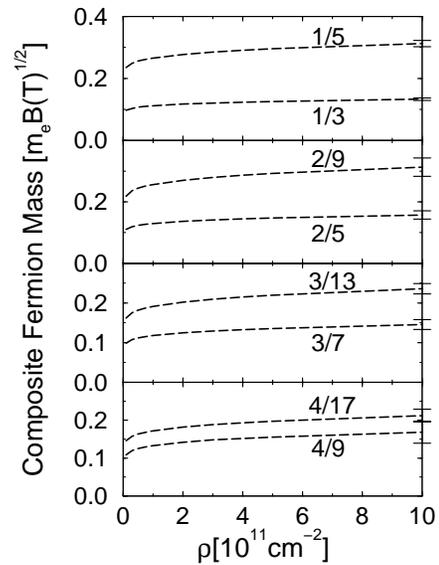,width=3.0in,angle=0}}
\caption{The masses of composite fermions carrying two and four flux quanta 
as a function of the two-dimensional density $\rho$, 
obtained from the gaps to charged excitations as discussed in the text.  The gaps
in turn were obtained from a thermodynamic
extrapolation of gaps for finite systems (up to 100 particles)
evaluated by Monte Carlo method. The
effective two dimensional interaction between electrons was 
calculated for a single heterojunction
in a local density approximation scheme.  There is no adjustable parameter in the
calculation.  The typical uncertainty coming from
the statistical sampling in Monte Carlo as well as from the extrapolation to the
thermodynamic limit is shown at the end of each curve. 
\label{fig1}}
\end{figure}

\begin{figure}
\centerline{\psfig{figure=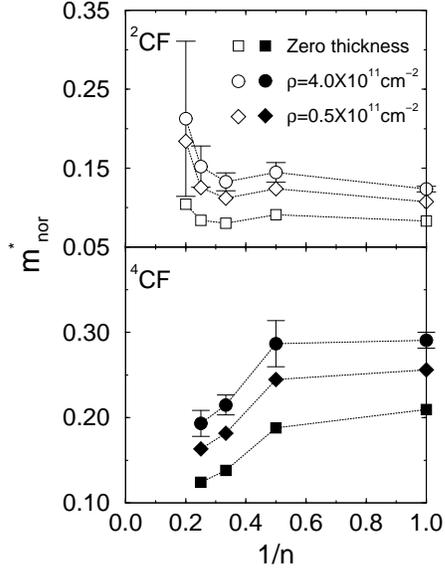,width=3.0in,angle=0}}
\caption{The $n$ dependence of the mass of composite fermion at 
$\nu={n}/{(2n+1)}$
and $\nu={n}/{(4n+1)}$ for two different densities with the effective
two-dimensional interaction evaluated in an LDA scheme, and also for 
a strictly two-dimensional system.  The mass is given 
in units of $m_e\sqrt{B[T]}$, where $m_e$ is
the electron mass in vacuum and $B$ is measured in Tesla.
The lines are a guide to the eye.
For clarity, the error bars are shown only for one density.
\label{fig2}}
\end{figure}

The gap to charged excitations was calculated  
for $\nu={1}/{5}$, ${2}/{9}$, ${3}/{13}$, and ${4}/{17}$ as a 
function of the density.  Fig.~(\ref{fig1})  shows the mass deduced from the 
gap, in units of $m_e\sqrt{B[T]}$,  and Fig.~(\ref{fig2}) shows the $n$  
dependence of the mass for certain typical densities.  The corresponding 
results for $^2$CFs are also shown in these figures for comparison \cite {Park99}.  
In the following, we discuss the conclusions that emerge from these results.

The reader may note that for a given 
filling factor, $m^*_{nor}$ is only very weakly dependent on density, 
as also found experimentally \cite {Yeh}.
For example, while the density changes by a factor of 20 
from $0.5\times 10^{11}$ cm$^{-2}$ to $1.0\times 10^{12}$ cm$^{-2}$, $m^*_{nor}$ 
changes only by $\sim$ 30\% or less; e.g., for $^2$CFs, it goes 
from approximately 0.12 to 0.13-0.16 for the filling factors shown. 
At first sight, 
this may seem precisely as expected, since for strictly zero thickness,
for which the interparticle interaction is proportional to 
${e^2}/{l_0}\sim \sqrt{B}$,  the factor $\sqrt{B}$ accounts for the 
entire density dependence of the mass (remember, the density and $B$ are 
related for a given filling factor).
However, in reality, there is another length scale controlled by the density, 
namely the width of the wave function in the direction
perpendicular to the plane, which also affects the interaction between 
electrons and gives a density dependence to $m^*_{nor}$.
To estimate the size of the correction, consider the widely used 
model potential\cite {ZDS} $V_{ZDS}(r)={(e^2/\epsilon)}(\lambda^2+r^2)^{-1/2}$.  The 
relevant quantity is the dimensionless `thickness' parameter $\lambda/l_0$, 
which changes by a factor of $\sim$ 4.5 over the density range stated above, 
assuming a constant $\lambda$;  for a change of $\lambda/l_0$ from 1.0 to 4.5, 
the results in Ref.~\onlinecite{Murthy} imply a change of $m^*_{nor}$ by a
factor of 4 to 6 at filling factors $\nu={1}/{3}$, 
${2}/{5}$ and ${3}/{7}$.  This underlines the importance of the use of 
a realistic interaction,  which leads to a much smaller variation of the normalized
mass as a function of 
the density.  Some insight into the issue of the density dependence can be gained 
from the Fang-Howard variational form for the transverse wave function \cite{Fang}, 
$\xi_{FH}\sim z\exp[-b z/2]$, where $z$ is the distance in the direction 
perpendicular to the plane
and $b\propto \rho^{1/3}$, neglecting the depletion charge density \cite{Fang}.  
The maximum of the wave function is obtained at $w=2/b$, which is a measure 
of the effective width in the transverse direction.  Noting that 
$l_0\sim \rho^{-1/2}$ for a fixed filling factor, we obtain for the 
dimensionless width, $w/l_0\propto \rho^{1/6}$; 
the smallness of the exponent gives an indication of why the 
density dependence of the mass is so weak.  
Coming back to $V_{ZDS}(r)$,  the above discussion implies that 
the relationship between $\lambda$ and the
``thickness" (or the density) is not straightforward for the heterojunction geometry.

A clear qualitative difference between the behaviors of the masses of 
$^2$CFs and $^4$CFs as a function of $n$, the number of filled CF-LLs, is 
manifest in Fig.~(\ref{fig2}), for the pure two-dimensional case as well as after 
including the finite thickness correction.  
The mass is seen to increase significantly 
in going from $\nu=4/17$ to $\nu=2/9$, in agreement with the 
experiment \cite{Pan}. 
What is the physical origin of this behavior?  By definition, for strictly
non-interacting composite fermions the $m^*_{nor}$ is a constant equal to the bare 
$m^*_{nor}$, implying that 
any variation in the mass originates from a renormalization of the bare mass 
by the residual interaction between composite fermions.  
One therefore expects to see a filling factor dependence of the CF mass 
only in regions where the inter-CF interaction is strong compared to 
the ``bare" gap.  One such region is 
in the vicinity of the composite fermion 
Fermi sea where the bare gap becomes rather small.  The interaction 
between composite fermions also becomes increasingly 
significant at small filling factors.
A recent work \cite{Park99b} has found that 
in the filling factor range ${1}/{4}>\nu>{1}/{5}$  
interaction between composite fermions is strong,
in fact comparable to the effective CF Fermi energy, causing a spontaneous
magnetization for $^4$CFs even when the Zeeman energy is negligible.
We believe that the same interaction between composite fermions causes 
the anomalous filling factor dependence of $m^*_{nor}$  in this region.  
This is indeed closely related to the physics attributed to 
this effect in Ref.~\onlinecite{Pan}, since the inter-CF 
interactions are also responsible for an instability into 
the Wigner crystal state at still lower filling factors \cite{JK}.  
Further, it is intuitively quite plausible that the mass 
is {\em enhanced}, rather than diminished,
by the inter-CF interaction, because it will likely 
broaden the CF-LLs and, leading to a gap reduction or a 
mass increment.

The Fig.~(\ref{fig2}) also gives a clue to how one may understand  
theoretically the experimentally discovered 
equality of the masses of $^2$CFs and $^4$CFs.
In the two-dimensional limit, the mass ratios 
$m^*_{nor}[\nu={n}/{(4n+1)}]/m^*_{nor}[\nu={n}/(2n+1)]$ is 
$\approx$ 2.5, 2.0, 1.7, and 1.4 for $n=1$, 2, 3, and 4, respectively.  One of the 
original motivations behind our calculations  was to investigate if the 
finite thickness correction might bring the ratios closer to unity,
but they turn out to be rather insensitive to the density. 
These ratios, however, are not really inconsistent with 
experiment, since the mass equality pertains to the masses {\em outside} 
the interaction dominated regions, i.e., to the bare masses, which raises the
question of how the bare masses can be deduced from the fully renormalized 
masses obtained in our calculations.  
For $^2$CFs, it is natural to fix the bare 
mass by  extrapolating the mass to $n\rightarrow \infty$ in Fig.~(\ref{fig2})
while leaving out the masses at $\nu={5}/{11}$ and $\nu={4}/{9}$, 
where the interaction effects become important.  
For $^4$CFs, it would be appropriate to determine masses from the states
$\nu={n}/{(4n-1)}$, following Refs.~\onlinecite {Yeh} and \onlinecite {Pan}, where
the filling factor dependence is small.  
However, these states are accessed from $n$ filled Landau levels by 
{\em reverse} flux attachment \cite{Jain,Wu}, the technical complications 
arising from which preclude us from carrying out calculations
directly at these filling factors.  We instead propose to determine the bare mass 
of $^4$CFs by performing an extrapolation of the masses at $\nu={2}/{9}$, 
$\nu={3}/{13}$ and
$\nu={4}/{17}$ to $\nu={1}/{4}$ in Fig.~(\ref{fig2}) \cite{Comment2}.
Quite remarkably, this procedure produces very similar values of $m^*_{nor}$'s for  
$^4$CFs and $^2$CFs, e.g., $m^*_{nor}\approx 0.12-0.13$ at $\rho=0.5\times 10^{11}$
cm$^{-2}$.  Furthermore, the bare $m^*_{nor}$ of
$^4$CFs is approximately one half of the $m^*_{nor}$
at $\nu={1}/{5}$ in our calculations.  In experiments, the interpretation of the
mass at $\nu={1}/{5}$ is complicated by the proximity to strong
localization; the mass at $\nu={4}/{5}$, which is $\nu={1}/{5}$ of holes,
may provide a more meaningful comparison to theory.
The mass at $\nu={4}/{5}$ was found\cite{Yeh} to be $m^*_{nor}=0.53-0.58$,
also approximately twice the value of the experimental bare mass
$m^*_{nor}=0.25-0.27$.  This general consistency with experiment 
provides an {\em a posteriori} justification for 
our method of obtaining the bare mass of $^4$CFs.

As mentioned earlier, there is no fundamental principle that forces the 
equality of the bare masses ... it just turns out to be so for the real situation.  
It is in fact straightforward to construct artificial potentials for which 
the bare masses are not equal.  For example, consider a model in which only two
pseudpotentials,\cite{Haldane} $V_1$ and $V_3$ are non-zero, with $V_1>>V_3.$ 
For $^2$CFs the gaps are determined by the full interaction, i.e., effectively 
by $V_1$. For $^4$CFs, a good approximation to the various eigenstates 
(exact in the limit ${V_3}/{V_1}\rightarrow 0$) is obtained by
multiplying the corresponding $^2$CF eigenstates by $\prod_{j<k}(z_j-z_k)^2$, 
where $z_j=x_j+i y_j$ is the coordinate of the $j$th electron.  This
eliminates states containing particles with relative angular momentum equal to unity, 
making eigenenergies independent of $V_1$.  Thus, in this model the  
mass of $^2$CFs is controlled by $V_1$ whereas that of $^4$CFs by $V_3$, and
consequently the two can be tuned independently of one another.

While we have obtained the qualitative behavior seen in 
experiment, the explanation is incomplete until the value of the 
experimental mass itself can be accounted for theoretically.  At the moment, the 
theoretical mass differs from the experimental one by roughly a factor of two.
A proper understanding of the discrepancy 
would ultimately require a quantitative 
theory of Shubnikov de Haas (SdH) oscillations of the composite
fermions, and of how the disorder affects the mass.
The deviation between the theoretical and 
experimental masses is at least consistent with what one might expect from 
disorder.  First, the  experimental bare mass for $^2$CFs and 
$^4$CFs ($m^*_{nor}\approx 0.25-0.27$) 
is bigger than the theoretical one ($m^*_{nor}\approx 0.12-0.13$).
Second, the values of the experimental and theoretical 
bare masses differ by roughly the {\em same} amount  (a factor of two) for 
{\em both} $^2$CFs and $^4$CFs.

In conclusion, the theory is able to reproduce certain {\em qualitative} 
features of the dependence of the composite fermion mass on various parameters. 
The following overall picture has emerged:  The bare masses of $^2$CFs and 
$^4$CFs are quite close, but are significantly renormalized 
in regions where the inter-CF interactions are non-negligible, 
specifically, close to the Fermi sea as well as for $\nu<{1}/{4}$.  
We thank V. Scarola, R. Shankar, and H. Stormer for useful discussions and 
comments, and gratefully acknowledge support by the
National Science Foundation under grant no. DMR-9615005,
and by the National Center for Supercomputing Applications at the University
of Illinois (Origin 2000), as well as 
the Numerically Intensive Computing Group at the Penn State University-CAC.

\begin{thebibliography}{99}


\bibitem{Yeh} A.S. Yeh, H.L. Stormer, D.C. Tsui, L.N. Pfeiffer, K.W. Baldwin, and
K.W. West, Phys. Rev. Lett. {\bf 82}, 592 (1999).

\bibitem{Pan} W. Pan, H.L. Stormer, D.C. Tsui, L.N. Pfeiffer, K.W. Baldwin, and
K.W. West, cond-mat/9910182.

\bibitem{Jain}  J.K. Jain, Phys. Rev. Lett. {\bf 63}, 199 (1989);
Phys. Rev. B {\bf 41}, 7653 (1990); J.K. Jain and R.K. Kamilla, in 
Ref.~\onlinecite {book1}.

\bibitem{book1} {\em Composite Fermions}, edited by 
Olle Heinonen (World Scientific, New Jersey, 1998).

\bibitem{book2} {\em Perspectives in Quantum Hall Effects}, edited by S. Das 
Sarma and A. Pinczuk (Wiley, New York, 1997).

\bibitem{Stormer} For a review, see 
H.L. Stormer and D.C. Tsui, in Ref.~\onlinecite{book2}.

\bibitem{HLR} B.I. Halperin, P.A. Lee, and N. Read, Phys. Rev. B {\bf
47}, 7312 (1993).

\bibitem{Park98} K. Park and J.K. Jain, Phys. Rev. Lett. {\bf 80}, 4237 (1998).

\bibitem{Melinte} 
S. Melinte, N. Freytag, M. Horvati\'c, C. Berthier, L.P. L\'evy, V.
Bayot, and M. Shayegan, cond-mat/9908098.

\bibitem{Kukushkin} I.V. Kukushkin, K. v. Klitzing, and K. Eberl, Phys. Rev.
Lett. {\bf 82}, 3665 (1999).

\bibitem{JK} J.K. Jain and R.K. Kamilla, Int. J. Mod. Phys. B {\bf 11},
2621 (1997); Phys. Rev. B {\bf 55}, R4895 (1996).

\bibitem{Bonesteel}  V. Melik-Alaverdian and N.E. Bonesteel, Phys. Rev. B {\bf
52}, R17032 (1995); Phys. Rev. Lett. {\bf 79}, 5286 (1997); R. Price and S. 
Das Sarma, Phys. Rev. B {\bf 52}, 17032 (1995).

\bibitem{Park99} K. Park, N. Meskini, and J.K. Jain, J. Phys. Condens. Matter {\bf
11}, 7283 (1999).

\bibitem{ZDS} F.C. Zhang and S. Das Sarma, Phys. Rev. B {\bf 33}, 2903 (1986). 

\bibitem{Murthy} G. Murthy {\em et al}, Phys. Rev. B {\bf 58}, 15363 (1998).

\bibitem{Fang} F.F. Fang and W.E. Howard, Phys. Rev. Lett. {\bf 16}, 797 (1966); F.
Stern and W.E. Howard, Phys. Ref. {\bf 163}, 816 (1967).

\bibitem{Park99b} K. Park and J.K. Jain, cond-mat/9910285.

\bibitem{Wu} X.G. Wu, G. Dev, and J.K. Jain, Phys. Rev. Lett. {\bf 71}, 153
(1993).

\bibitem{Comment2} It is possible that the mass may start {\em 
increasing} closer to the $\nu={1}/{4}$ Fermi sea; such an increase 
is obviously of no relevance to our extrapolation aimed at  
extracting the {\em bare} mass. 

\bibitem{Haldane} The pseudopotential $V_m$ is the energy of two electrons in a state
of relative angular momentum $m$.  Any interaction is fully specified by 
the values of $V_m$ in lowest Landau level.  For further details, see, F.D.M. 
Haldane, Phys. Rev.  Lett. {\bf 51}, 605 (1983).


\end{thebibliography}
\end{document}